\documentclass[prl,twocolumn,showpacs,superscriptaddress]{revtex4}

\usepackage{bm}
\usepackage{graphicx}
\usepackage{amsmath}
\usepackage{amssymb}

\def\be{\begin{equation}}
\def\ee{\end{equation}}

\begin{document}

\begin{abstract}
This is an extended Reply to Comment by A. Sergeev, M.Y. Reizer, and V. Mitin
[arXiv:0906.2389] on our Letter [Phys. Rev. Lett. \textbf{102}, 067001 (2009)].
We explicitly demonstrate that all claims by Sergeev et al. are completely
unfounded, because their underlying theoretical work contains multiple
errors and inconsistencies. For this reason, there is no need to revise
the existing theories of thermoelectric response in superconductors.
\end{abstract}

\title{On the Nernst effect in fluctuating superconductors: \\
Serbyn, Skvortsov, and Varlamov reply}
\author{M. N. Serbyn}
\affiliation{Department of Physics, Massachusetts Institute of Technology,
Cambridge, Massachusetts 02139, USA}
\affiliation{Landau Institute for Theoretical Physics, Chernogolovka, Moscow Region,
142432, Russia}
\author{M. A. Skvortsov}
\affiliation{Landau Institute for Theoretical Physics, Chernogolovka, Moscow Region,
142432, Russia}
\author{A. A. Varlamov}
\affiliation{SPIN-CNR, Viale del Politecnico 1, I-00133 Rome, Italy}
\date{\today}
\maketitle



\section{Introduction}

In a series of recent papers (see, Ref.~\cite{SRM-comment} and references
therein), Sergeev, Reizer, and Mitin (SRM) have argued that all existing
results on thermoelectric response in fluctuating superconductors are
fundamentally flawed and must be revisited. According to SRM, these previous
allegedly incorrect works include the pioneering works of Ullah and Dorsey~%
\cite{Dorsey} and those of Ussishkin, Sondhi, and Huse~\cite{Uss1,Uss2,Uss3}%
, the relevant chapter in the book of Varlamov and Larkin~\cite{LV}, our
recent Letter~\cite{we}, and the related work of Michaeli and Finkelstein~%
\cite{Fin}. Note that these researches~\cite{Dorsey,Uss1,Uss2,Uss3,LV,we,Fin}
have employed at least four different approaches to calculate the Nernst
effect in a superconductor: Time-dependent Ginzburg-Landau theory~\cite%
{Uss1,Uss3,LV}, microscopic techniques based on the Matsubara diagrammatic
technique~\cite{Uss2,we} and on the quantum kinetic equation~\cite{Fin}, and
finally simple phenomenological arguments, which relate the Nernst
coefficient to the temperature dependence of the chemical potential for the
carriers~\cite{we}. All these approaches consistently provide the same order
of magnitude for the Nernst coefficient, which has been shown to much exceed
that in a Fermi liquid (hence, we called the effect ``giant'' in our
Letter). Such a giant Nernst signal has been detected in a variety of
well-known experiments in the high-temperature cuprates \cite{Ong} and
conventional superconducting films \cite{Aubin1,Aubin2}. Therefore the
existence of giant Nernst effect constitutes a solid experimental fact,
which has been the actual motivation for the aforementioned theoretical
works. However, SRM ``strongly object'' to the existence of these results
and experimental facts, arguing~\cite{SRM-PRB} that the \textquotedblleft
numerous recent theories grossly overestimate the thermomagnetic
coefficients\textquotedblright\ and calling them in the abstract of their
recent comment~\cite{SRM-comment} \textquotedblleft ridiculously
large.\textquotedblright\ These strong claims of Sergeev \emph{et al.}\/ are
based solely on their own alternative calculation approach, sketched by them
in Ref.~\cite{SRM-PRB}.

In 2009 SRM had escalated the concern about the existing theories of the
Nernst effect, by posting a comment \cite{SRM-comment} on our Letter.
Despite SRM's evident errors, their Comment has recently been accepted
for publication in Physical Review Letters. This manuscript
is an extended version of our Reply.

Our Reply contains the following: (i)~In the first part, we analyze the
paper~\cite{SRM-PRB}, which is needed because all SRM's criticism on the
existing theories is based on this single paper. Hence, a careful
analysis of Ref.~\cite{SRM-PRB} is the only means for us to refute SRM's
criticism of the works by us and others. As a result of this exercise, we
are able to show in the first part of the Reply that Ref.~\cite{SRM-PRB} contain serious errors. We
identify the main problems of SRM's treatment as likely originating from the
combination of inconsistent use of a specific gauge in a single calculation,
missing pieces in the relevant diagrams, and technical mistakes.
(ii)~Next, we suggest specific technical steps to remedy these problems of
SRM and bring their approach in accordance with the exiting theories. But
most importantly, we conclude that since the paper \cite{SRM-PRB} is
manifestly incorrect, there is no need to revise all previous existing
theories. (iii)~In the last part, we elaborate on the phenomenological
Eq.~(1) that we suggested in our Letter \cite{we} as a simple intuitive
argument behind the experimentally observed giant Nernst effect.

\section{Analysis of Phys. Rev. B \textbf{77}, 064501 (2008) by Sergeev et
al.}

Ref.~\cite{SRM-PRB} concentrates entirely on the discussion of ``a gauge-invariant
microscopic approach,'' but we show below that it is the very
gauge-invariance that is explicitly violated in the SRM's treatment. Below
we choose to focus on two specific technical problems (among many) that
suffice to prove the inadequacy of their proposed calculational method and
therefore make the criticism mute.

\subsection{On the calculations of the Nernst effect in a normal metal by
Sergeev et al.}

One of only a few existing and commonly-accepted results on the Nernst
effect is that of Sondheimer~\cite{Sond}, who first calculated the effect in
a Fermi gas back in 1948. Any technique that claims to be of relevance to
more complicated Nernst phenomena, such as those due to superconducting
fluctuations, must recover Sondheimer's formula as a basic \textquotedblleft
sanity check\textquotedblright. In Sec.~II of their work, SRM do consider
the case of a normal metal but fail to explain the origin of Sondheimer's~\cite{Sond}
formula in a comprehensible way. In addition,
their discussion contains an error, which propagates
into further analysis.

For noninteracting fermions, the thermoelectric response is described by the
Kubo-like diagram constructed of the heat and electric vertices and two
Green functions. In the absence of a magnetic field, the heat vertex is $\xi_%
\mathbf{p}\mathbf{v} = \xi_\mathbf{p} \mathbf{p}/m$. For nonzero magnetic field, both
the kinetic energy $\xi_\mathbf{p}$ and the momentum $\mathbf{p}$ itself,
should be modified by the vector potential $\mathbf{A}$. SRM forget about
the latter and thus come to a \emph{wrong expression} for the heat current
(Eq.~(5) of Ref.~\cite{SRM-PRB}):
\begin{equation}
\mathbf{J}^h_\text{SRM} = \sum_{\mathbf{p}} \mathbf{v}\xi_{\mathbf{p}}a_{%
\mathbf{p}}^+a_{\mathbf{p}} + \sum_{\mathbf{p}} \frac{e\mathbf{v}}{c} (%
\mathbf{v}\cdot \mathbf{A}) a_{\mathbf{p}}^+a_{\mathbf{p}} .  \label{SRM-5}
\end{equation}
Further they forget to incorporate the vector potential $\mathbf{A}$ into
the electric current to maintain gauge invariance. Then the correlator $%
\langle j^{Q}j^{e}\rangle $ is claimed to be calculated, but (i)~a {gauge in
which calculations are performed is poorly defined} and (ii)~{the external
momentum is unspecified} (below, we explain why it is important). We
reiterate here that while SRM concentrate their work as well as their
criticism of others on the gauge-invariance issue, which indeed is
important, they fail to satisfy the gauge invariance in their own work, even
at the non-interacting level. Their \textquotedblleft
gauge-invariant\textquotedblright\ expressions contain explicitly a vector
potential, which is inserted into selected pieces of diagrams in an
uncontrolled fashion and calculations are performed in an unspecified gauge.

\subsection{On the calculations of the Nernst effect in a fluctuating
superconductor by Sergeev et al.}

In Sec.~II of Ref.~\cite{SRM-PRB} Sergeev \emph{et al.}\ analyze the heat
current transferred by fluctuating Cooper pairs in the vicinity of $T_{c}$.
They correctly obtain that the heat current is proportional to the
gauge-invariant momentum (Eq.~(11) of Ref.~\cite{SRM-PRB}):
\begin{equation}
\mathbf{B}_{\text{SRM}}^{h}\propto \omega (\mathbf{q}+2e\mathbf{A}/c),
\label{SRM-11}
\end{equation}%
and they claim that it is the term with $\mathbf{A}$ that is their key new
finding and that this term was allegedly overlooked in all previous
calculations. However, SRM again fail to include $\mathbf{A}$ in the electric
vertex and draw the diagrams, extracting $\mathbf{A}$ from the propagators
and from the heat vertex (\ref{SRM-11})
only. Then they say: \textquotedblleft Further calculations of the diagrams
for Cooper pairs {are similar} to that for noninteracting
electrons\textquotedblright, thus inheriting all the inconsistencies from
their normal-metal treatment. Even without that, \emph{the expressions
determining current} (\ref{SRM-5}) and (\ref{SRM-11}) \emph{are different
in their vector structure}: the first one contains
$\mathbf{v}(\mathbf{v}\cdot\mathbf{A})$, while the second does not!

\subsection{Correcting the problem with gauge invariance}

Here we elaborate on the main technical problems of SRM's calculations and
provide technical details on how to correctly enforce gauge-invariance for
observables. Of course, thermal conductivity and any other linear response
for a physical observable must be gauge-invariant independently of the means
used to calculate it. In the framework of Kubo diagrammatic approach (both
for a normal metal and a superconductor close to $T_{c}$), the relevant
diagrams for thermoelectric response contain three sources of the vector
potential/magnetic-field dependence: (A)~Green functions/propagators,
(B)~heat current vertex, and (C)~electric current vertex. The resulting
expression is gauge invariant only if all these three sources are taken into
account in a consistent fashion and within a specific gauge. In Ref.~\cite%
{SRM-PRB} it is claimed that the contribution (B) cancels the contribution
(A) calculated by Ussishkin in Ref.~\cite{Uss2}. However the consistency
between calculations and gauge choices in the two parts of the same physical
quantity is not discussed. Most importantly, contribution (C) is not even
mentioned by Sergeev et al., which makes their conclusion erroneous. This
contrasts sharply with our approach, where we follow the work of Galitski
and Larkin~\cite{VGL}, and from the outset use Landau representation that
automatically takes into account all sorts of couplings to the magnetic
field in a manifestly gauge-invariant fashion.

We have found that the use of Landau basis greatly simplifies the
calculations and automatically circumvents the need to keep track of gauge
invariance at intermediate steps. However, near $T_{c}$, the use of this
method is not required and correct result can be obtained via perturbative
expansions of all elements of the diagram in the vector potential. In this
approach, care is required to keep the gauge invariance under control at all
stages of the calculation. The correct starting point is to consider the
correlator \setcounter{equation}{0}
\begin{equation}
\Pi_{\alpha\beta}(\mathbf{k},\mathbf{Q}) = \langle j_\alpha^Q(\mathbf{k}+%
\mathbf{Q})j_\beta^e(-\mathbf{k})\rangle ,
\end{equation}
where $\mathbf{Q}$ is the momentum associated with the vector potential: $%
\mathbf{A}(\mathbf{r})=\mathbf{a}\,e^{i\mathbf{Q}\cdot \mathbf{r}}$. Setting
$\mathbf{k}=0$ implies that the momentum $\mathbf{Q}$ comes into the heat
vertex, while setting $\mathbf{k}=-\mathbf{Q}$ corresponds to $\mathbf{Q}$
coming out of the electric vertex. The limit $\mathbf{k}\rightarrow 0$ and $%
\mathbf{Q}\rightarrow 0$ needs to be taken in the end.

Now we expand all terms (two vertices and two propagators) to linear order
in $\mathbf{A}$. In the most general form, this procedure generates four
contributions $\Pi^{(m)}$ ($m=1,\dots,4$) of the form (we retained the
leading order in $\mathbf{k}$ and $\mathbf{Q}$):
\begin{equation}
\Pi_{\alpha \beta }^{(m)} = \mathcal{S}_{\alpha \beta \gamma
\mu}^{(m)}k_{\gamma }a_{\mu } + \mathcal{T}_{\alpha \beta \gamma
\mu}^{(m)}Q_{\gamma }a_{\mu }.
\end{equation}
Since $\mathcal{S}_{\alpha \beta \gamma \mu }^{(m)}\neq 0$, the relative
contribution of these four terms retains the value of $\mathbf{k}$. This is
a reason why it is important to specify $\mathbf{k}$ when calculating each
of the four gauge-noninvariant pieces. Of course, $\sum_{m}\mathcal{S}%
_{\alpha \beta \gamma \mu }^{(m)}=0$, indicating that there is a
well-defined limit $\Pi _{\alpha \beta }(0,\mathbf{Q})$. The contributions $%
\mathcal{T}_{\alpha\beta\gamma\mu}^{(m)}$ are generally non-zero and
explicitly gauge-dependent, but their sum is certainly gauge-invariant:
\begin{equation}
\Pi _{\alpha \beta } \propto (Q_{\alpha }a_{\beta }-Q_{\beta }a_{\alpha})
\int d^dq\,q^{2}[L_{+}^{2}L_{-}-L_{+}L_{-}^{2}] ,
\end{equation}
where $L_+=L(i\varepsilon_n+i\omega_\nu,q)$ and $L_-=L(i\varepsilon_n,q)$
are the fluctuation propagators. After summing over the Matsubara
frequencies, $\varepsilon_n$, and performing an analytic continuation in $%
\omega_\nu$ we come to the following result for the off-diagonal
thermomagnetic response:
\begin{equation}
\tilde\beta_\text{C.p.}^{xy} \propto \int d^{d}q\,q^{2}\!\int d\omega
\,\omega \frac{\partial B(\omega )}{\partial \omega }|L(\omega ,q)|^{2} %
\mathop{\rm Im} L(\omega,q) ,  \label{beta-s}
\end{equation}
where $B(\omega)=\coth(\omega/2T)$ is the bosonic equilibrium distribution
function, and
\begin{equation}
L(\omega,q) = -\frac{8T_c/\pi\nu}{(8/\pi)(T-T_c) - i\omega + Dq^2}
\end{equation}
is the retarded propagator for Cooper pairs. Evaluating the integrals in
Eq.~(\ref{beta-s}), one immediately finds a finite value for Nernst
coefficient $\propto(T-T_c)^{d/2-2}$ even in the absence of the
particle-hole asymmetry, in accordance with all previous works in the field
\cite{Dorsey,Uss1,Uss2,Uss3,LV,we,Fin}.

\subsection{Thermomagnetic response: bosons vs.\ fermions}

To clarify the relation between the thermomagnetic responses of fluctuating
Cooper pairs in the vicinity of $T_c$ [Eq.~(\ref{beta-s})] and that of
normal electrons, $\beta^{xy}_{n}$, we find it instructive to present the
latter in a similar form:
\begin{equation}  \label{beta-n}
\beta^{xy}_{n} \propto \int d^dp \, p^2 \! \int d\varepsilon \, \varepsilon
\, \frac{\partial F(\varepsilon)}{\partial\varepsilon} \,
|G(\varepsilon,p)|^2 \mathop{\rm Im} G(\varepsilon,p) ,
\end{equation}
where $F(\varepsilon)=\tanh(\varepsilon/2T)$ is the fermionic equilibrium
distribution function, and
\begin{equation}
G(\varepsilon,p) = \frac{1}{\varepsilon - \xi_\mathbf{p} + i/2\tau}
\end{equation}
is the retarded electron Green function. The similarly looking expressions (%
\ref{beta-s}) and (\ref{beta-n}) lead to essentially different results for
the Nernst coefficient for electrons and Cooper pairs. For free fermions in
the absence of the particle-hole asymmetry, momentum integration in Eq.~(\ref%
{beta-n}) gives an $\varepsilon$-independent constant, and the subsequent
energy integral vanishes by oddness. On the other hand, for fluctuating
Cooper pairs the integrals in Eq.~(\ref{beta-s}) lead to a finite Nernst
coefficient even in the absence of the particle-hole asymmetry. The same
conclusion can be readily achieved within the time-dependent Ginzburg-Landau
approach \cite{LV}.

\subsection{Role of magnetization}


Sergeev et al.~\cite{SRM-comment} claim that the contribution of
magnetization is not relevant for calculating the Nernst coefficient. This
statement evidently contradicts the well established theory of
thermomagnetic effects (see, e.g., \cite{ZG}). According to the theory, the
experimentally measured value of the thermoelectric tensor $%
\beta_{\alpha\beta}$ is given by a sum of the kinetic contribution, $%
\tilde\beta_{\alpha\beta}$, (which can be calculated using the Kubo-like
approach, see above) and a thermodynamic contribution due to magnetization $%
\mathbf{M}$:
\begin{equation}
\beta^{\alpha\beta} = \tilde\beta^{\alpha\beta} +
\epsilon_{\alpha\beta\gamma} c M_\gamma/T .  \label{beta-sum}
\end{equation}
The importance of the magnetization heat current for the Nernst effect was
first demonstrated by Obraztsov in 1965 \cite{Obraztsov}. Later, the
significance of this contribution to thermomagnetic response of electron
systems has been emphasized by a number of authors in relation to the
integer quantized Hall effect, interacting electron gas in a quantizing
magnetic field \cite{CHR97}, and the fluctuation Cooper pairs contribution
to the Nernst effect within Ginzburg-Landau formalism \cite{Uss1,LV}. The
crucial role of the magnetization contribution to the Nernst effect has
recently been demonstrated in Refs.~\cite{we,Fin} where it cancels the
otherwise divergent kinetic contribution $\tilde\beta_{xy}$ at low
temperatures above $H_{c2}$, thus being eventually responsible for the
implementation of the third law of thermodynamics.

SRM argued that magnetization effects do not contribute to the heat
transport in a magnetic field, and the second term in Eq.~(\ref{beta-sum})
should be omitted. This statement contradicts all known theories and is
incorrect. As a side remark, we mention here that even if the magnetization
contribution is omitted, the Nernst effect in fluctuating superconductors
would still be giant provided that the error in calculating $\tilde\beta_{xy}
$ made in Ref.~\cite{SRM-PRB} is corrected.

Finally, it also should be mentioned that the splitting of the Nernst
coefficient (\ref{beta-sum}) in a kinetic (Kubo-like) and thermodynamic
(magnetization) contribution is completely analogous to that in the Hall
conductivity: $\sigma_{xy}=\sigma _{xy}^{\text{I}}+\sigma _{xy}^{\text{II}}$
\cite{sigmaxy}.

\subsection{Summary}

To summarize this part: pretending to have developed a gauge-invariant
microscopic approach to heat transfer, the article \cite{SRM-PRB} starts
with a manifestly gauge non-invariant expression for the heat current
[Eq.~(5)] even for free electrons. It is precisely the incomplete account
for the gauge invariance for why the authors of Ref.~\cite{SRM-PRB} came to
the wrong conclusion that the nonzero Nernst effect due to fluctuation
Cooper pairs necessarily requires presence of the particle-hole asymmetry in
the underlying electronic band structure in a metallic phase.

\section{On the Qualitative formula for the Nernst coefficient}

We take the opportunity here to discuss in more detail the qualitative
expression for the Nernst coefficient,
\begin{equation}
N = \frac{\sigma}{nq^{2}c} \, \frac{d\mu}{dT} ,  \label{nernst}
\end{equation}
which was suggested in our Letter~\cite{we} and that was also criticized by
SRM~\cite{SRM-comment}.

We note that the main purpose of our work was to present a technically
challenging microscopic calculation of the Nernst coefficient for an
extended range of magnetic fields and temperatures. These microscopic
calculations do not rely on any phenomenology such as Eq.~(\ref{nernst}),
but are strongly supported by it. The main purpose of including Eq.~(\ref%
{nernst}) in our paper was to provide a simple intuitive argument behind the
giant Nernst effect in a fluctuating superconductor, which may be of value
to the wide audience of PRL. And we believe that phenomenological Eq.~(\ref%
{nernst}) is a new interesting (qualitative) result that does convey the
desired message.

\subsection{Application to fluctuating Cooper pairs}

Arguing for the validity of \emph{phenomenological}\/ Eq.~(\ref{nernst}) we
used the notion of a drift velocity and ignored impurity scattering of the
charge carriers, keeping in mind that the fluctuating Cooper pairs are not
scattered by elastic impurities (all information about such scattering is
included in the value of the effective coherence length or equivalently in
the effective mass of the Cooper pairs). Equation (\ref{nernst}) clarifies
the physical origin of the observed anomaly and, moreover, sheds light on
where to look for the giant Nernst effect: in systems with chemical
potential strongly dependent on temperature.

Now we clarify the issue related to the chemical potential of fluctuating
Cooper pairs, $\mu _{\text{C.p.}}(T)$, which we introduced as an auxiliary
concept within the phenomenological derivation of the Nernst effect. Indeed,
it is known that in the thermodynamic equilibrium, the chemical potential of
a system with a variable number of particles is zero, with photon and phonon
gases being the textbook examples. A na\"{\i}ve application of this
``theorem'' to fluctuating Cooper pairs ``gas'' has lead SRM to
a wrong conclusion that $\mu _{\text{C.p.}}=0$ \cite{SRM-comment}.
However, a delicate issue concerning Cooper pairs is that
they do not form an isolated system but are composed of fermion
quasiparticles which
constitute another subsystem under consideration. According to the same
textbook discussion \cite{mu-book}, in a multicomponent system, the chemical
potential of the $i$'th component, $\mu _{i}$, can be defined as the
derivative of the thermodynamic potential with respect to the number of
particles of $i$-th sort:
\begin{equation}
\mu_i = \left( \partial \Omega/\partial N_{i}\right)_{P,V,N_{j}} ,
\label{term}
\end{equation}
provided the numbers of particles of all other species are fixed, $N_{j\neq
i}=\text{const}$. In deriving the condition for thermodynamic equilibrium,
one should now take into account that creation of a Cooper pair must be
accompanied by removing two quasiparticles from the fermionic subsystem.
This leads to $\mu_\text{C.p.}-2\mu_{n}=0$, where $\mu _{n}$ is the chemical
potential of quasiparticles. Therefore,
the equilibrium condition does not restrict $\mu _{\text{C.p.}}$ to zero,
even though the number of Cooper pairs is not conserved.

The value of $\mu_\text{C.p.}(T)=T-T_c$ \cite{we} can be found from Eq.~(\ref%
{term}) using the explicit temperature dependencies of fluctuation part of
Gibbs potential $\Omega_\text{fl}$ and concentration of fluctuating Cooper
pairs. After this Eq.~(\ref{nernst}) reproduces \cite{we} the known results
for fluctuating Nernst response near the classical superconducting
transition point. This style of reasoning is in close analogy with the
``pedestrian'' approach to fluctuating paraconductivity~\cite{LV}, which
employs the standard Drude formula with the assumption that the
Ginzburg-Landau relaxation time plays the role of the scattering time. In
this sense, Eq.~(\ref{nernst}) is an analogue of the classical Drude formula
for electrical conductivity. This analogy also suggests that the domain of
validity of Eq.~(\ref{nernst}) is constrained to temperatures close to $%
T_{c}.$ In the opposite low-temperature limit, only microscopic calculations
(such as presented by us in Ref.~\cite{we}) are reliable.

What concerns the applicability of Eq.~(\ref{nernst}) to a normal metal,
one can easily check that it readily reproduces the Sondheimer's result.

\section{Conclusions}

To summarize, we list below certain points of the criticism of Ref.~\cite%
{SRM-comment} together with our responses:

\begin{enumerate}
\item SRM claim that ``the linear response calculation of $\beta_{xy}$ does
not require any magnetization correction'' and present their own
interpretation of various contribution to the heat current. --- We certainly
disagree with this statement. In such a delicate and controversial issue as
heat transport, we consider counterproductive to discuss interpretations.
Instead, one should either microscopically derive the expression for $%
\beta_{\alpha\beta}$ or check existing ones for possible inconsistencies.
The crucial role of the magnetization contribution to $\beta_{xy}$ has
recently been demonstrated in Refs.~\cite{we,Fin}: In particular, at $T\to0$
it cancels the otherwise divergent $\tilde\beta_{xy}$, thus ensuring the
implementation of the third law of thermodynamics. Therefore, omitting the
magnetization contribution to $\beta_{xy}$ as suggested by SRM will
inevitably violate the fundamental law of thermodynamics.

\item SRM claim that the particle-hole asymmetry of the single-electron
spectrum is necessary for a nonzero Nernst effect due to fluctuating Cooper
pairs. --- This wrong statement is solely based on the results of Ref.~\cite%
{SRM-PRB} which contains multiple errors, as discussed in details above.

\item SRM claim that our results \cite{we} generalized to repulsive
interaction in the Cooper channel would yield $\beta_{xy}$ significantly
exceeding that for non-interacting electrons. Further they claim that,
``certainly, this effect is not known''. --- We believe that the first of
these statements is correct and the Cooper-channel correction to $\beta_{xy}$
is indeed nonzero in the absence of the particle-hole asymmetry. Similar
effects have been theoretically predicted for fluctuation diamagnetism \cite%
{Maki73, Bulaevski74, AslamazovLarkin74} and fluctuation conductivity \cite%
{AltshulerReizerVarlamov83}. However we disagree with the second statement
of SRM: The relative values of various contributions to $\beta_{xy}$ should
be considered individually for each experiment.

\item SRM claim that, ``according to textbooks \cite{Ziman}, $\nabla\mu$
should always be included in the effective electric field''. --- That is
true, the textbook condition of vanishing current, $E-\nabla \mu /e=0$, is
precisely our initial assumption of electroneutrality employed in the
phenomenological treatment.

\item SRM claim that ``a thermodynamic value of $\mu_\text{C.p.}$ is always
zero, because a number of pairs is not conserved''. --- This statement is
wrong: The chemical potential of fluctuating Cooper pairs is nonzero in
equilibrium since they do not form an isolated system.
\end{enumerate}

Thus, it has been explicitly demonstrated that all SRM's ground-breaking
claims, seemingly of fundamental importance, are completely unfounded,
because the underlying theoretical work of SRM contains~\cite{SRM-PRB}
multiple errors and inconsistencies. For this reason, there is no need to
revise the existing theories of thermoelectric response in superconductors
and in particular results of a microscopic analysis presented in our Letter~%
\cite{we}; they remain intact.


\end{document}